# Volatility Depends on Market Trade and Macro Theory

Victor Olkhov
Independent, Moscow, Russia
victor.olkhov@gmail.com
ORCID iD 0000-0003-0944-5113

**June 14, 2024**

## Abstract

We consider the randomness of market trade as the origin of price and return stochasticity. We look at time series of trade values and volumes as random variables during the averaging interval $\Delta$ and describe the dependences of market-based volatilities of price and return on the volatilities and correlations of market trade values and volumes. We describe the market-based origin of the lower boundaries of the accuracy of macroeconomic variables and consider, as an example, the accuracy of macroeconomic investments. We highlight that current macroeconomic models describe relations between the 1st order variables determined by sums of trade values or volumes. To predict market-based volatilities of price, return, and volatilities of macroeconomic variables, one should develop econometric methodologies, collect data, and elaborate macroeconomic theories of the 2nd order that model the mutual dependence of the 1st and 2nd order economic variables. The absence of macroeconomic theories of the 2nd order means no economic basis for predictions of market-based volatilities of price and return, as well as volatilities of any macroeconomic variables. In turn, that limits the accuracy of forecasting probabilities of price, return, and the accuracy of macroeconomic variables in the best case by Gaussian distributions.

---

This research received no support, specific grants, or financial assistance from funding agencies in the public, commercial, or nonprofit sectors. We welcome valuable offers of grants, support, and positions.

# 1. Introduction

The predictions of price and return volatility have been under research for decades. Price behavior was studied by Muth (1961), Fama (1965), Stigler and Kindahl (1970), Friedman (1990), Cochrane (2001), Cochrane and Culp (2003), Nakamura and Steinsson (2008), Borovička and Hansen (2012), Weyl (2019), and many others. Price and return volatility and price-volume correlation were studied by Tauchen and Pitts (1983), Mankiw, Romer, and Shapiro (1991), Campbell, Grossman, and Wang (1993), Ito and Lin, (1993), Brock and LeBaron (1995), Bernanke and Gertler (1999), Shiryaev (1999), Andersen et al. (2001), Plerou et al. (2001), Poon and Granger (2003), Andersen et al. (2005), Ciner and Sackley (2007), Miloudi, Bouattour, and Benkraiem (2016). These references are only a small part of the countless studies.

Volatility describes the uncertainty of price, return, and the uncertainty of macroeconomic variables. The contribution of our paper to the volatility puzzle is the description of the dependence of the volatilities of price, return, and macroeconomic investment, as an example of macroeconomic variables, on random properties of market trade. We consider that all factors that can induce stochasticity of price and return are already imprinted in the randomness of market trade values $C(t_i)$ and volumes $U(t_i)$, and the randomness of price $p(t_i)$ at time $t_i$ is described by the trade price equation:

$$C(t_i) = p(t_i)U(t_i) \tag{1.1}$$

We assume that the time interval $\varepsilon$ between market trades is rather small and constant. We consider the time series of trade values $C(t_i)$ and volumes $U(t_i)$ (1.1) as random variables during a particular time averaging interval $\Delta$. The different choice of $\Delta$ results in different values of statistical moments of trade values and volumes. We consider (1.1) as the basis for the definition of market-based statistical moments of price and return and derive the dependence of market-based volatilities of price and return on volatilities and the correlation of trade values and volumes during $\Delta$.

In Section 2, we consider statistical moments of market trade values and volumes. We use the well-known volume weighted average price (VWAP) (Berkowitz et al., 1989; Duffie and Dworczak, 2018) as a market-based average and derive the direct dependence of market-based volatility of price on volatilities and the correlation of trade values and volumes. We highlight that different definitions of price averaging result in different properties of price statistical moments. In particular, we show that the use of VWAP results in zero price-volume correlation. In Section 3, we consider market-based volatility of return. As a market-based average return, we consider Markowitz's definition of portfolio return given in his famous paper (Markowitz, 1952). We discuss the similarities between the definitions of portfolio return and VWAP and derive the market-based volatility of return that depends on volatilities and the correlation between current and past trade values in a way similar to the dependence of market-based volatility of price. In Section 4, we outline the randomness of market trade as the root of the obstacles that impede accurate forecasts of the probability of price and return. We highlight market trade randomness as the cause of lower limitations of the accuracy of macroeconomic variables. As an example, we consider macroeconomic investment and describe the dependence of its volatility and the coefficient of variation on volatility and the coefficient of variation of market trade values. We show, that the predictions of volatilities of price and return require the development of macroeconomic theory that should describe macroeconomic variables composed of sums of squares of trade values and volumes. Conclusion in Section 5.

## 2. Market-based volatility of price

We start with the choice (2.1) of the time averaging interval $\Delta$:

$$t-\frac{\Delta}{2}< t_i < t+\frac{\Delta}{2} \quad ; \quad i=1,\dots N \tag{2.1}$$

We assume that all prices $p(t_i)$ are adjusted to the current time $t$. We consider the time series of market trade values $C(t_i)$ and volumes $U(t_i)$ as random variables during $\Delta$ (2.1). We define the n-th statistical moments of trade value $C(t;n)$ and volume $U(t;n)$:

$$C(t;n)=E[C^n(t_i)]\sim\frac{1}{N}\sum_{i=1}^{N}C^n(t_i) \quad ; \quad U(t;n)=E[U^n(t_i)]\sim\frac{1}{N}\sum_{i=1}^{N}U^n(t_i) \tag{2.2}$$

In (2.2), we denote mathematical expectation $E[..]$ and use the symbol "~" to highlight that relations (2.2) define estimates of statistical moments of random variables by a finite number $N$ of terms. We denote sums $C_\Delta(t;n)$ of the *n-th* degrees of trade values $C^n(t_i)$ and sums $U_\Delta(t;n)$ of the *n-th* degrees of volumes $U^n(t_i)$ during $\Delta$ (2.1) as (2.3):

$$C_\Delta(t;n)=N\cdot C(t;n)=\sum_{i=1}^{N}C^n(t_i) \quad ; \quad U_\Delta(t;n)=N\cdot U(t;n)=\sum_{i=1}^{N}C^n(t_i) \tag{2.3}$$

One can express VWAP $p(t;1,1)$ (Berkowitz et al., 1989; Duffie and Dworczak, 2018) as:

$$p(t;1,1)=\frac{1}{U_\Delta(t;1)}\sum_{i=1}^{N}p(t_i)U(t_i)=\frac{C_\Delta(t;1)}{U_\Delta(t;1)}=\frac{C(t;1)}{U(t;1)} \tag{2.4}$$

To highlight the meaning of VWAP (2.4) we consider the n-the degrees of (1.1):

$$C^n(t_i)=p^n(t_i)U^n(t_i) \tag{2.5}$$

The use of (2.5) for each degree $n=1,2,.$ and the relations (2.2-2.4) permit define the *m-th* statistical moments $p(t;m,n)$ of price in a form that is similar to (2.4):

$$p(t;m,n)=\frac{1}{U_\Delta(t;n)}\sum_{i=1}^{N}p^m(t_i)U^n(t_i) \quad ; \quad p(t;n,n)=\frac{C_\Delta(t;n)}{U_\Delta(t;n)}=\frac{C(t;n)}{U(t;n)} \tag{2.6}$$

One can present relations (2.6) in a different form:

$$p(t;m,n) = \sum_{i=1}^{N} p^m(t_i)\, w(t,t_i;n) \quad ; \quad w(t,t_i;n) = \frac{U^n(t_i)}{U_\Delta(t;n)} \quad ; \quad \sum_{i=1}^{N} w(t,t_i;n) = 1 \qquad (2.7)$$

Functions $w(t,t_i;n)$ (2.7) have the meaning of weight functions but not price probabilities. To derive the probability $\mu(p;t,n)$ of price $p$ one should sum all trade volumes that correspond to market trades at price $p$. The probability $\mu(p;t,n)$ of price $p$ takes the form:

$$\mu(p;t,n) = \frac{1}{U_\Delta(t;n)} \sum_{p(t_i)=p} U^n(t_i) \quad ; \quad p(t;m,n) = \sum_p p^m\, \mu(p;t,n) \qquad (2.8)$$

The sum in (2.8) is taken over all values of market prices $p$ during the interval $\Delta$. However, the estimates (2.7) of weight functions $w(t,t_i;n)$ are sufficient to assess statistical moments of price $p(t;m,n)$ that correspond to equation (2.5) as the *n-th* degree of (1.1).

We consider VWAP $p(t;1,1)$ (2.4) as market-based average price $a(t;1)$ and define:

$$a(t;1) = E_m[p(t_i)] = p(t;1,1) \quad ; \quad a(t;n) = E_m[p^n(t_i)] \qquad (2.9)$$

We use $E_m[..]$ to denote market-based mathematical expectation of price and market-based *n*-th statistical moments $a(t;n)$ (2.9) of price to distinguish from $E[..]$ (2.2), which we note as frequency-based mathematical expectation, and from price statistical moments $p(t;m,n)$ determined by weight functions $w(t,t_i;n)$ (2.7).

Equations (2.5) and relations (2.6) define the *n-th* statistical moment of price $p(t;n,n)$ in a form that is alike to the form of VWAP. In other words, the *n-th* statistical moment of price $p(t;n,n)$ is determined as the ratio of the total sum of *n-th* degrees of trade values $C_\Delta(t;n)$ during $\Delta$ to the total sum of *n-th* degrees of trade volumes $U_\Delta(t;n)$. That is the economic essence of the average of the *n-th* degree of price during $\Delta$. However, each next *n-th* statistical moment of price $p(t;n,n)$ for $n=1,2,3,..$ is determined by different weight functions $w(t,t_i;n)$ (2.7), and hence can be non-consistent with the previous ones. To derive a self-consistent set of market-based statistical moments of price starting with market-based average price $a(t;1)$ (2.9), one should reconcile each next price statistical moment $p(t;n,n)$ with the previous ones.

To derive the 2nd market-based statistical moment $a(t;2)$ (2.9) and market-based price volatility $\sigma^2(t)$ (2.10), one should prove that market-based volatility of price $\sigma^2(t)$ (2.10) is non-negative and $a(t;2) \geq a^2(t;1)$:

$$\sigma^2(t) = E_m\left[\big(p(t_i) - a(t;1)\big)^2\right] = a(t;2) - a^2(t;1) \geq 0 \qquad (2.10)$$

To satisfy the condition (2.10) on the 2nd market-based statistical moment $a(t;2)$ of price, we determine price volatility $\sigma^2(t)$ (2.11) through the 2nd weight functions $w(t,t_i;2)$ (2.7):

$$\sigma^2(t) = \sum_{i=1}^{N} \big(p(t_i) - a(t;1)\big)^2 w(t,t_i;2) = p(t;2,2) - 2p(t;1,2)a(t;1) + a^2(t;1) \qquad (2.11)$$

It is obvious, that (2.11) satisfies non-negative condition. From (2.10; 2.11) obtain

$$a(t;2) = p(t;2,2) + 2a(t;1)[a(t;1) - p(t;1,2)] \qquad (2.12)$$

In App.A, we derive the dependence of volatility $\sigma^2(t)$ (2.13) on volatilities (2.15) of trade values $\Omega_C^{\ 2}(t)$ and volumes $\Omega_U^{\ 2}(t)$ and on their correlation (2.16):

$$\sigma^2(t) = \frac{\Omega_C^2(t) + a^2(t;1)\Omega_U^2(t) - 2a(t;1)corr\{C(t)U(t)\}}{U(t;2)} \qquad (2.13)$$

From (2.10; 2.13) obtain expression for $a(t;2)$ (2.14):

$$a(t;2) = E_m[p^2(t_i)] = \frac{C(t;2) + 2a^2(t;1)\Omega_U^2(t) - 2a(t;1)corr\{C(t)U(t)\}}{U(t;2)} \qquad (2.14)$$

We define trade value volatility $\Omega_C^{\ 2}(t)$ and trade volume volatility $\Omega_U^{\ 2}(t)$ (2.15):

$$\Omega_C^2(t) = C(t;2) - C^2(t;1) \quad ; \quad \Omega_U^2(t) = U(t;2) - U^2(t;1) \qquad (2.15)$$

The correlation $corr\{C(t)U(t)\}$ (2.16) between trade value and volume during $\Delta$:

$$corr\{C(t)U(t)\} = E\big[\big(C(t_i) - C(t;1)\big)\big(U(t_i) - U(t;1)\big)\big] = E[C(t_i)U(t_i)] - C(t;1)U(t;1) \qquad (2.16)$$

The joint average $E[C(t_i)U(t_i)]$ (2.17) of the product of trade value and volume takes the form:

$$E[C(t_i)U(t_i)] = \frac{1}{N}\sum_{i=1}^{N} C(t_i)U(t_i) \qquad (2.17)$$

The relations (2.13) highlight the dependence of the market-based volatility $\sigma^2(t)$ of price on the volatility of trade value $\Omega_C^{\ 2}(t)$, volatility of trade volume $\Omega_U^{\ 2}(t)$ (2.15) and correlation $corr\{C(t)U(t)\}$ (2.16) between trade value and volume during $\Delta$. The volatility $\sigma^2(t)$ (2.13) depends also on the 2nd statistical moment of trade volume $U(t;2)$ and on the market-based average price $a(t;1)$ (2.9), which is equal to VWAP $p(t;1,1)$ (2.4). That completely describes the dependence of market-based volatility $\sigma^2(t)$ (2.13) of price during $\Delta$ on the statistical properties of random market trade values and volumes.

The use of VWAP $p(t;1,1)$ as market-based average price results in no price-volume correlation. Indeed, from (1.1; 2.2) and the definition of VWAP $p(t;1,1)$ (2.4), one obtains:

$$E[C(t_i)] = C(t;1) = E[p(t_i)U(t_i)] = p(t;1,1)U(t;1)$$

Hence, the market-based price-volume correlation $corr\{p(t)U(t)\}$ is always zero:

$$corr\{p(t)U(t)\} = E[p(t_i)U(t_i)] - p(t;1,1)U(t;1) = 0 \qquad (2.18)$$

Actually, the price-volume correlations were studied in numerous papers (Karpoff, 1987; Gallant, Rossi, and Tauchen, 1992; Miloudi, Bouattour, and Benkraiem, 2016). The authors present evidence of positive or negative price-volume correlations. However, these papers assess price-volume correlations using frequency-based estimates of price that are different from the estimates made under a market-based approach. In particular, the estimates of statistical moments $\pi(t;n)$ of price $p(t_i)$ during the interval $\Delta$ are similar to the assessments of statistical moments $U(t;n)$ (2.2) of trade volumes:

$$P(p) \sim \frac{m(p)}{N} \quad ; \quad \pi(t;n) = E[p^n(t_i)] \sim \frac{1}{N}\sum_{i=1}^{N} p^n(t_i) \tag{2.19}$$

In (2.19) *m(p)* denotes number of trades at price *p*. It is obvious, that (2.6; 2.7) and market-based statistical moments of price *a(t;1)* (2.9) and *a(t;2)* (2.14) differ from statistical moments *π(t;1)* and *π(t;2)* (2.19). Statistical moments of price *π(t;n)* (2.19) don't take into account the impact of random properties of market trade volumes. That is the main and most important distinction between assessments of frequency-based and market-based statistical moments of price. One can easily show that relations (2.6; 2.7), expressions for market-based average *a(t;1)* (2.9), and price volatility $\sigma^2(t)$ (2.13) coincide with frequency-based definitions (2.19) if all trade volumes $U(t_i)$ are constant during the averaging interval *Δ*. The different approaches to the assessments of price statistical moments result in different conclusions on price-volume correlation. We refer to Olkhov (2021a; 2022) for further details and derivation of the 3rd and 4th market-based statistical moments of price.

## 3. Market-based volatility of return

The description of random properties of return, modeling, and predictions of volatility of return are studied by many authors (Fama, 1965; Brock and LeBaron, 1995; Andersen et al., 2001; 2005; Poon and Granger, 2003). These researchers consider the random properties of return and the volatility of return using frequency-based considerations of return time series. In simple words, the probability *P(r)* (3.1) of return *r* is assumed to be proportional to the number of terms *m(r)* of return *r,* similar to (2.19), and frequency-based *n-th* statistical moments of return *ρ(t;n)* take the form (3.1):

$$P(r) \sim \frac{m(r)}{N} \quad ; \quad \rho(t;n) = E[r^n(t_i)] \sim \frac{1}{N}\sum_{i=1}^{N} r^n(t_i) \tag{3.1}$$

However, in his famous paper on portfolio choice, Markowitz (1952) proposed a definition of portfolio return *r(t,τ;1,1)* (3.6) that has a form almost identical to the form of VWAP (2.4) and differs from the frequency-based average return *ρ(t;1)* (3.1). We use Markowitz's definition as a definition of market-based average return and introduce market-based volatility of return in a way similar to the description of market-based volatility of price. We consider return $r(t_i,\tau)$ (3.2) as a ratio of price $p(t_i)$ at time $t_i$ to price $p(t_i-\tau)$ in the past at $t_i-\tau$:

$$r(t_i,\tau) = \frac{p(t_i)}{p(t_i-\tau)} \tag{3.2}$$

We assume that $t_i$ for *i=1,..N* belongs to the averaging interval *Δ* (2.1) and estimate the average and volatility of return during the interval *Δ* (2.1) with respect to the constant time shift *τ*. To do that, let us transform the trade price equation (1.1) as follows:

$$C(t_i) = p(t_i)U(t_i) = \frac{p(t_i)}{p(t_i-\tau)}\, p(t_i-\tau)U(t_i) = r(t_i,\tau)\, C_o(t_i,\tau) \tag{3.3}$$

In (3.3), we denote the past value $C_o(t_i,\tau)$ (3.4) of the current trade volume $U(t_i)$ at price $p(t_i-\tau)$ at time at $t_i-\tau$ in the past:

$$C_o(t_i,\tau)=p(t_i-\tau)U(t_i) \tag{3.4}$$

Equation (3.5) has a form similar to (1.1), and thus we can use the same considerations to describe the market-based average and volatility of return (3.1).

$$C(t_i)=r(t_i,\tau)\;C_o(t_i,\tau) \tag{3.5}$$

Similar to (1.1), from equation (3.5), it follows that the statistical properties of the current $C(t_i)$ and past $C_o(t_i,\tau)$ trade values completely determine the statistical properties of market returns $r(t_i,\tau)$. For $t_i$, $i=1,\ldots N$, during the averaging interval $\Delta$ (2.1), one can consider returns $r(t_i,\tau)$ (3.2) as returns (3.5) on assets of the portfolio with current trade values $C(t_i)$ and past values $C_o(t_i,\tau)$. Let us define the *n-th* statistical moments $C_o(t,\tau;n)$ of past value $C_o(t_i,\tau)$ and total past trade value $C_{o\Delta}(t,\tau;n)$ during $\Delta$ (2.1) similar to (2.2; 2.3):

$$C_o(t,\tau;n)=\frac{1}{N}\sum_{i=1}^{N}C_o^n(t_i,\tau) \quad ; \quad C_{o\Delta}(t,\tau;n)=N\cdot C_o(t,\tau;n)=\sum_{i=1}^{N}C_o^n(t_i,\tau) \tag{3.6}$$

Then, Markowitz's (1952) definition of portfolio return $r(t,\tau;1,1)$ weighted by their past values, takes the form (3.7; 3.8), similar to the form of VWAP (2.4; 2.7):

$$r(t,\tau;1{,}1)=\frac{1}{C_{o\Delta}(t,\tau;n)}\sum_{i=1}^{N}r(t_i,\tau)\;C_o(t_i,\tau)=\frac{C_\Delta(t;1)}{C_{o\Delta}(t,\tau;n)}=\frac{C(t;1)}{C_o(t,\tau;n)} \tag{3.7}$$

$$r(t,\tau;1{,}1)=\sum_{i=1}^{N}r(t_i,\tau)z(t_i,\tau;1) \tag{3.8}$$

Relations (3.8) present the average return $r(t,\tau;1,1)$ as returns weighted by their relative past values $z(t,t_i,\tau;1)$ (3.9) during the averaging interval $\Delta$ (2.1):

$$z(t,t_i,\tau;1)=\frac{C_o(t_i,\tau)}{\sum_{i=1}^{N}C_o(t_i,\tau)}=\frac{C_o(t_i,\tau)}{C_{o\Delta}(t,\tau;n)} \quad ; \quad \sum_{i=1}^{N}z(t,t_i,\tau;1)=1 \tag{3.9}$$

We take relations (3.7; 3.8) as the market-based average return $h(t,\tau;1)$:

$$h(t,\tau;1)=E_m[r(t_i,\tau)]=r(t,\tau;1{,}1) \tag{3.10}$$

Similar to (2.9), we denote $E_m[..]$ (3.10) market-based mathematical expectation of return to distinguish it from frequency-based averaging (3.1). To define the market-based volatility of return $\varphi^2(t,\tau)$ (3.13; 3.16), we repeat the steps for derivation of the market-based volatility $\sigma^2(t)$ (2.13) of price. We consider the *n-th* degree of (3.5):

$$C^n(t_i)=r^n(t_i,\tau)\;C_o^n(t_i,\tau) \tag{3.11}$$

Similar to (2.6; 2.7), we define the *m-th* statistical moments of return $r(t,\tau;m,n)$ (3.12) averaged over weight functions $z(t,t_i,\tau;n)$ (3.13):

$$r(t,\tau;m,n)=\frac{1}{C_{o\Delta}(t,\tau;n)}\sum_{i=1}^{N}r^m(t_i,\tau)\;C_o^n(t_i,\tau) \;;\quad r(t,\tau;n,n)=\frac{C_\Delta(t;n)}{C_{o\Delta}(t,\tau;n)}=\frac{C(t;n)}{C_o(t,\tau;n)} \tag{3.12}$$

$$r(t,\tau;m,n)=\sum_{i=1}^{N}r^m(t_i,\tau)z(t,t_i,\tau;n) \;;\; z(t,t_i,\tau;n)=\frac{C_o^n(t_i,\tau)}{C_{o\Delta}(t,\tau;n)} \;;\; \sum_{i=1}^{N}z(t,t_i,\tau;n)=1 \tag{3.13}$$

The functions $z(t,t_i,\tau;n)$ (3.13) have the meaning of weight functions but not probabilities of return, similar to the functions $w(t,t_i;n)$ (2.7). For $n=1,2,\ldots$ weigh functions $z(t,t_i,\tau;n)$ (3.13) define *m-th* statistical moments of return $r(t,\tau;m,n)$ (3.12).

To define the market-based 2nd statistical moment $h(t,\tau;2)$ and volatility $\varphi^2(t,\tau)$ of return that are approved with market-based average return $h(t,\tau;1)$, we follow the derivation (2.10-2.17). To reconcile the market-based 2nd statistical moment of return $h(t,\tau;2)$ that should be determined by the weight functions $z(t,t_i,\tau;2)$ (3.13) with the average return $h(t,\tau;1)$ that is determined by the weight functions $z(t,t_i,\tau;1)$ (3.9), we define the market-based volatility $\varphi^2(t,\tau)$ of return:

$$\varphi^2(t,\tau) = E_m\left[\left(r(t_i,\tau) - h(t,\tau;1)\right)^2\right] = h(t,\tau;2) - h^2(t,\tau;1) \geq 0 \qquad (3.14)$$

To fulfill (3.14), we determine market-based volatility $\varphi^2(t,\tau)$ of return (3.15) as:

$$\varphi^2(t,\tau) = \sum_{i=1}^{N}\left(r(t_i,\tau) - h(t,\tau;1)\right)^2 z(t,t_i,\tau;2) = r(t,\tau;2,2) - 2r(t,\tau;1,2)h(t,\tau;1) + h^2(t,\tau;1) \qquad (3.15)$$

For the 2nd statistical moment $h(t,\tau;2)$ of return obtain:

$$h(t,\tau;2) = r(t,\tau;2,2) + 2h(t,\tau;1)[h(t,\tau;1) - r(t,\tau;1,2)] \qquad (3.16)$$

Similar to (2.13; 2.14) obtain:

$$\varphi^2(t,\tau) = \frac{\Omega_C^2(t) + h^2(t,\tau;1)\Omega_{Co}^2(t,\tau) - 2h(t,\tau;1)corr\{C(t)C_o(t,\tau)\}}{C_o(t,\tau;2)} \qquad (3.17)$$

$$h(t,\tau;2) = \frac{C(t;2) + 2h^2(t,\tau;1)\Omega_{Co}^2(t,\tau) - 2h(t,\tau;1)corr\{C(t)C_o(t,\tau)\}}{C_o(t,\tau;2)} \qquad (3.18)$$

In (3.17; 3.18) we use (3.6) to denote volatility $\Omega_{Co}^{\ 2}(t,\tau)$ (3.19) of past values $C_o(t_i,\tau)$:

$$\Omega_{Co}^2(t,\tau) = C_o(t,\tau;2) - C_o^2(t,\tau;1) \qquad (3.19)$$

The correlation $corr\{C(t)C_o(t,\tau)\}$ (3.20) between current $C(t)$ and past $C_o(t,\tau)$ trade values during the averaging interval $\Delta$ (2.1) takes the form:

$$corr\{C(t)C_o(t,\tau)\} = E\left[\left(C(t_i) - C(t;1)\right)\left(C_o(t_i,\tau) - C_o(t,\tau;1)\right)\right]$$

$$corr\{C(t)C_o(t,\tau)\} = E[C(t_i)C_o(t_i,\tau)] - C(t;1)C_o(t,\tau;1) \qquad (3.20)$$

The joint average $E[C(t_i)C_o(t_i,\tau)]$ (3.21) of the product of current $C(t_i)$ and past $C_o(t_i,\tau)$ trade values takes the form:

$$E[C(t_i)C_o(t_i,\tau)] = \frac{1}{N}\sum_{i=1}^{N} C(t_i)C_o(t_i,\tau) \qquad (3.21)$$

The relations (3.14-3.21) determine the dependence of the market-based 2nd statistical moment $h(t,\tau;2)$ (3.18) and volatility $\varphi^2(t,\tau)$ (3.17) of return on volatilities of current $\Omega_C^{\ 2}(t)$ (2.15) and past $\Omega_{Co}^{\ 2}(t,\tau)$ (3.19) trade values and on their mutual correlation $corr\{C(t)C_o(t,\tau)\}$ (3.20). We refer to Olkhov (2023a) for further details.

## 4. Volatility as a Piece of Macro Financial Puzzle

The direct dependences of market-based volatilities of price and return on volatilities and correlations of market trade (2.13-2.16; 3.17-3.20) during the averaging interval $\Delta$ (2.1) uncover the origin of economic-based limitations of the accuracy of the forecasts of price and return probabilities, as well as the limitations of the accuracy of predictions of probabilities of economic and financial variables at horizon $T$.

One can consider the uncertainty of the economic and market environment, the action of numerous risks, the unpredictability of agents' expectations, and many other factors as the origin of market trade randomness. However, the action of all these factors results in a random time series of market trade values and volumes during the averaging interval $\Delta$. To simplify the complex description of direct and backward influences that impact market trade, we consider random market trade time series as the main, initial origin of random economic evolution and development and as the origin of price and return stochasticity. We consider the random change of economic and financial variables as the result of random market trades, transactions, and deals between agents.

We regard trade values, volumes, price, and return as random variables during the averaging interval $\Delta$ (2.1). The predictions of random variables at horizon $T$ mean the predictions of their random properties at the same horizon. The properties of a random variable can be described equally by a probability measure, a characteristic function, or a set of statistical moments (Shiryaev, 1999; Shreve, 2004). A finite number of statistical moments describe the approximations of the characteristic function and probability of a random variable. The more statistical moments of a random variable that could be predicted, the higher the accuracy of the predictions that could be obtained. In particular, the predictions of the first two statistical moments describe the Gaussian approximation of the probability of a random variable.

The dependences of the market-based volatilities of price (2.13) and return (3.17) on the volatilities and correlation of market trade values and volumes uncover the origin of economic-based limitations of the accuracy of forecasts of probabilities of price and return by Gaussian distributions. Indeed, the uncertainty of predictions of market-based average price $a(t;1)$ (2.9) or VWAP price $p(t;1,1)$ (2.4) is described by the forecasts of price volatility $\sigma^2(t)$ (2.13). To assess the accuracy of predictions of the average price $a(t;1)$ (2.9) at horizon $T$, one should forecast the price volatility $\sigma^2(t)$ (2.13) at the same horizon. However, to do that, one should forecast volatilities (2.15) and correlations (2.16) of trade values and volumes at the same horizon $T$. And that raises tough problems.

### *4.1. The limitations of the accuracy of macroeconomic variables*

The limitations of the accuracy of macroeconomic variables are consequences of the direct and hidden dependence of macroeconomic variables on random values and volumes of market trade. The lack of direct initial data of market trade values and volumes that are made by agents that are required for the direct assessment of the change of macroeconomic variables during a particular time averaging interval requires the development and use of econometric methodologies that estimate macroeconomic variables using available econometric data. Different methodologies for the econometric assessments of economic variables may vary a bit and give different interpretations. For certainty, we choose Fox et al. (2019) as the perfect methodology for assessing macroeconomic variables and refer to it for any details. The changes of *additive* economic and financial variables, such as investment and credits, consumption and production, etc., are determined by sums of linear combinations of the products of trade values or volumes and particular coefficients during the averaging interval $\Delta$. *Non-additive* variables, such as prices, inflation, growth rates, etc., are determined as ratios of *additive* macro variables. We highlight that almost all variables are determined as sums of linear combinations of trade values or volumes with deterministic or random coefficients, and we consider these linear combinations as a random variable during $\Delta$.

We outline the important statement: the accuracy of assessments and forecasts of macroeconomic variables is determined by two different factors. The first factor defines the lower bound of accuracy of the value of the macroeconomic variable that is determined by the volatility of corresponding market trade values or volumes during $\Delta$. The second factor determines the accuracy of the approximations of macroeconomic variables according to the econometric methodology and the use of the available economic data. In this paper, we consider the first factor.

For simplicity, as an example, we consider the assessment of the uncertainty of macroeconomic investment during the interval $\Delta$. The economic sense of macroeconomic investment is determined by the sum of investments made by all economic agents during $\Delta$. In turn, investments made by a particular economic agent are determined by the sum of all investment transactions made by that agent during $\Delta$. The absence of the initial investment transactions made by all agents during $\Delta$ requires the use of econometric methodologies for the assessment of macroeconomic investment. The use of indirect data that is different from agents' investment transactions adds excess uncertainty to the estimates of investment.

The uncertainty of macroeconomic investment is determined by the random nature (2.2; 2.3; 2.15) of market trade values during *Δ*. The interval *Δ* can be equal to a week, month, quarter, or whatever. Let us define $C(t_i;j)$ investment transaction that is made by agent $j$ at time $t_i$ and assume that each agent $j$, $j=1,..M$, during *Δ* (2.1) made $i=1,..N$ investment transactions, so the total number of investment transactions equals $N{\cdot}M$. We assume that the investment transactions $C(t_i;j)$ don't repeat or duplicate each other. With use of (2.3), one can define macroeconomic investment $In(t;1)$ during *Δ* as:

$$In(t;1) = \sum_{j=1}^{M}\sum_{i=1}^{N} C(t_i;j) = C_\Delta(t;1) \tag{4.1}$$

The macroeconomic investment $In(t;1)$ (4.1) during *Δ* is determined by sum of random variables $C(t_i;j)$ and that defines the origin of its uncertainty. To explain and derive the uncertainty of macroeconomic investment $In(t;1)$ (4.1) made during the averaging interval *Δ* we define an "instantaneous" macroeconomic investment $In(t_i;j)$ (4.2):

$$In(t_i;j) = N \cdot M \cdot C(t_i;j) \tag{4.2}$$

The "instantaneous" investment $In(t_i;j)$ (4.2) is determined by $N{\cdot}M$ random investment transactions $C(t_i;j)$. From (2.2; 2.3), obtain that the average $E[In(t_i;j)]$ (4.3) of the "instantaneous" investment $In(t_i;j)$ (4.2) equals macro investment $In(t;1)$ (4.1) during *Δ:*

$$E[In(t_i;j)] = N \cdot M \cdot E[\, C(t_i;j)] = N \cdot M \cdot C(t;1) = C_\Delta(t;1) = In(t;1) \tag{4.3}$$

$$C_\Delta(t;1) = \sum_{j=1}^{M}\sum_{i=1}^{N} C(t_i;j) = N \cdot M \cdot C(t;1) \; ; \; C(t;1) = \frac{1}{N \cdot M}\sum_{j=1}^{M}\sum_{i=1}^{N} C(t_i;j) \tag{4.4}$$

One can consider the random "instantaneous" investment $In(t_i;j)$ (4.2) as the origin of macroeconomic investment $In(t;1)$ (4.1) during *Δ*. Hence, one can consider the volatility $\Omega_{In}^2(t)$ (4.5; 4.7) of the "instantaneous" investment $In(t_i;j)$ (4.2) as the uncertainty of macroeconomic investment $In(t;1)$ (4.1) during *Δ*. From (2.2; 2.15; 4.2; 4.3), obtain:

$$\Omega_{In}^2(t) = E[(In(t_i;j) - In(t;1))^2] = E[In^2(t_i;j)] - In^2(t;1) \tag{4.5}$$

$$E[In^2(t_i;j)] = (N \cdot M)^2 E[C^2(t_i;j)] = (N \cdot M)^2 C(t;2) \tag{4.6}$$

$$\Omega_{In}^2(t) = (N \cdot M)^2 [C(t;2) - C^2(t;1)] = (N \cdot M)^2\, \Omega_C^2(t) \tag{4.7}$$

From (4.1; 4.3; 4.7), obtain that the square of the coefficient of variation $\chi_{In}^2(t)$ (4.8) of macroeconomic investment equals the coefficient of variation $\chi_C^2(t)$ of investment transactions during *Δ*:

$$\chi_{In}^2(t) = \frac{\Omega_{In}^2(t)}{In^2(t;1)} = \frac{\Omega_C^2(t)}{C^2(t;1)} = \chi_C^2(t) \tag{4.8}$$

We underline that the volatility $\Omega_{In}^2(t)$ (4.5) and the coefficient of variation $\chi_{In}^2(t)$ (4.8) of the "instantaneous" investment $In(t_i;j)$ (4.2) depend on the sum of squares of investment transactions $In(t;2)$ (4.9) during *Δ:*

$$In(t;2) = \sum_{j=1}^{M}\sum_{i=1}^{N} C^2(t_i;j) = C_\Delta(t;2) = N \cdot M\, C(t;2) = \frac{1}{N \cdot M} E[In^2(t_i;j)] \quad (4.9)$$

The 2nd statistical moment of "instantaneous" investment equals:

$$E[In^2(t_i;j)] = N \cdot M\, In(t;2) \quad (4.10)$$

The average $In(t;1)$ (4.1; 4.3) of the "instantaneous" investments $In(t_i;j)$ (4.2) and its 2nd statistical moment (4.10) define the Gaussian approximation of the probability of the "instantaneous" macroeconomic investment $In(t_i;j)$. The forecasts of $In(t;1)$ (4.1; 4.3) and $N \cdot M\, In(t;2)$ (4.9; 4.10) define the volatility $\Omega_{In}^2(t)$ (4.5; 4.7) and the coefficient of variation $\chi_{In}^2(t)$ (4.8) of the "instantaneous" investment $In(t_i;j)$ (4.2) at horizon T. Almost equal assessments result from the predictions of $C(t;1)$, $C(t;2)$ (2.2), and volatility $\Omega_C^2(t)$ (2.15) at horizon $T$. These predictions define the lower bound of the accuracy of macroeconomic investment at horizon T. On the other hand, they determine the forecasts of the Gaussian approximation of the probability of the "instantaneous" investment $In(t_i;j)$ (4.2) at horizon $T$.

All these forecasts are linked with predictions of the 1st and 2nd statistical moments of market trade values. Current macroeconomic theories describe relations between the 1st order economic variables determined by sums of trade values and volumes. To forecast economic variables that depend on squares of trade values and volumes, one should develop macroeconomic theories of the 2nd order that describe relations between the 1st and 2nd order variables composed of sums of the squares of trade values and volumes (Olkhov, 2021b; 2023b; 2023c).

We underline the main issue: the lower economic boundaries of the accuracy of macroeconomic investment (4.5; 4.7) or other macro variables is determined by the volatilities (2.15) or the coefficient of variation $\chi_C^2(t)$ (4.8) of market trades.

The 2nd order "twins", similar to $In(t;2)$ (4.9; 4.10) should be calculated for other additive macroeconomic variables of the 1st order. The 2nd order variables compose the set of macroeconomic variables that can't be expressed or described through variables of the 1st order. The set of 1st and 2nd order economic variables establishes the basis for the development of macroeconomic theory of the 2nd order. That problem at least doubles the complexity of macroeconomic theory. In the case of success, macroeconomic theory of the 2nd order will provide economic-based predictions of the volatilities (2.15; 3.19) and correlations (2.16; 3.20) that are required for forecasting market-based volatilities of price and return (2.13; 3.17). We highlight that volatilities define only the 2nd statistical moments. Hence, the long and complex route that should be taken with creating the methodologies for the assessments of the 2nd order economic variables, and the development of the 2nd order macroeconomic theories will give us only economic-based predictions of the 2nd statistical moments of price, return, and other economic variables. That will help us to predict probabilities of price and return only in Gaussian approximations. Till that, the current predictions of Gaussian approximations have almost no economic foundation.

To forecast probabilities of price or return with accuracy beyond Gaussian approximation, one should establish econometric methodologies, collect data, and develop theories that describe economic variables of the 3rd and higher orders that are composed of 3rd and higher degrees of trade values or volumes. Those are good aims for the long future.

## 5. Conclusion

Actually, all economic and financial variables are subject to volatility. The randomness of market trade values and volumes, the stochasticity of price and return, and the randomness of macroeconomic variables make studies of volatility an essential and mandatory part of theoretical economics. Obviously, investigations of volatilities of price and return are the most "beneficial" subjects for investors, and predictions of random properties of price and return will remain in demand for investors, economic modeling, and forecasting for a long time. We present a market-based consideration of the volatilities of price and return. Our approach results in strict limitations of the accuracy of predictions of price and return probabilities, the limitations of the accuracy of macroeconomic variables, and highlight the relations between random properties of market trade and macroeconomic theory.

However, investors are free to make trading decisions that are based on their own expectations and forecasts of random properties of price and return and have nothing in common with the results described above. The collisions between market-based random properties of price and return and unpredictable investors' expectations that result in trade decisions add immense complexity to studies that could develop macroeconomic theories of the 2nd order and create an economic basis for predictions of volatilities and correlations of trade values and volumes. These and many other obstacles for many years to come securely limit the accuracy of forecasts of price and return probabilities in the best case by Gaussian approximations.

# Appendix

## Market-based volatility of price

Let us use (2.2; 2.3; 2.5) and present price statistical moments *p(t;m,n)* (2.7) as follows:

$$p(t;m,n)U(t;n) = \frac{1}{N}\sum_{i=1}^{N} p^m(t_i)U^n(t_i) = \frac{1}{N}\sum_{i=1}^{N} C^m(t_i)U^{n-m}(t_i) = CU(t;m,n-m) \quad (A.1)$$

The joint average of m-th degree of trade value $C^m(t_i)$ and the (n-m)-th degree o trade volume $U^{n-m}(t_i)$ (A.1) can be presented as:

$$CU(t;m,n-m) = C(t;m)U(t;n-m) + corr\{C^m(t)U^{n-m}(t)\} \quad (A.2)$$

From relations (A.1; A.2) we transform (2.11) as:

$$C(t;2) = p(t;2,2)U(t;2) \;\; ; \quad p(t;1,2)U(t;2) = C(t;1)U(t;1) + corr\{C(t)U(t)\}$$

Using (2.6; 2.9) one can present price volatility (2.11) as:

$$\sigma^2(t) = p(t;2,2) - 2p(t;1,2)a(t;1) + a^2(t;1)$$

$$= \frac{C(t;2) - 2(C(t;1)U(t;1) + corr\{C(t)U(t)\})a(t;1)}{U(t;2)} + a^2(t;1)$$

$$\sigma^2(t) = \frac{C(t;2) - 2C^2(t;1) - 2a(t;1)corr\{C(t)U(t)\}}{U(t;2)} + a^2(t;1)$$

$$\sigma^2(t) = \frac{\Omega_C^2(t) - C^2(t;1) - 2a(t;1)corr\{C(t)U(t)\}}{U(t;2)} + a^2(t;1)$$

Volatility $\Omega_C^2(t)$ of trade value takes form (2.15). Then from (2.4; 2.9) obtain:

$$a^2(t;1) - \frac{C^2(t;1)}{U(t;2)} = a^2(t;1)\left[1 - \frac{U^2(t;1)}{U(t;2)}\right] = \frac{a^2(t;1)\Omega_U^2(t)}{U(t;2)}$$

Hence, obtain expression for price volatility (2.13):

$$\sigma^2(t) = \frac{\Omega_C^2(t) + a^2(t;1)\Omega_U^2(t) - 2a(t;1)corr\{C(t)U(t)\}}{U(t;2)}$$

## References

Andersen, T., Bollerslev, T., Diebold, F.X. and H. Ebens, (2001). The Distribution of Realized Stock Return Volatility, Journal of Financial Economics, 61, 43-76

Andersen, T.G., Bollerslev, T., Christoffersen, P.F. and F.X. Diebold, (2005). Volatility Forecasting, CFS WP 2005/08, 1-116

Berkowitz, S.A., Dennis E. Logue, D.E. and E. A. Noser, Jr., (1988). The Total Cost of Transactions on the NYSE, The Journal Of Finance, 43, (1), 97-112

Bernanke, B. and M. Gertler, (1999). Monetary Policy and Asset Price Volatility. FRB of Kansas City, Economic Review, 4Q, 1-36

Borovička, J. and L. P. Hansen, 2012. Examining Macroeconomic Models through the Lens of Asset Pricing. FRB Chicago

Brock, W.A. and B.D. LeBaron, (1995). A Dynamic structural model for stock return volatility and trading volume. NBER, WP 4988, 1-46

Campbell, J.Y., Grossman, S.J., Wang, J., 1993. Trading Volume And Serial Correlation In Stock Returns. The Quarterly Journal of Economics, 905-939

Ciner, C. and W. H. Sackley, 2007. Transactions, volume and volatility: evidence from an emerging market, Applied Financial Economics Letters, 3, 161-164

Cochrane, J.H., 2001. Asset Pricing. Princeton Univ. Press, Princeton, US

Cochrane, J.H. and C.L. Culp, 2003. Equilibrium Asset Pricing and Discount Factors: Overview and Implications for Derivatives Valuation and Risk Management. In Modern Risk Management. A History, Ed. S.Jenkins, 57-92

Duffie, D. and P. Dworczak, (2018). Robust Benchmark Design, NBER WP 20540, 1-56

Fama, E.F., 1965. The Behavior of Stock-Market Prices. The Journal of Business, 38, (1), 34-105.

Fox, D.R., et.al. (2019). Concepts and Methods of the U.S. National Income and Product Accounts. BEA, Dep.Commerce, US, Chapters 1-13, 1- 449

Friedman, D.D., 1990. Price Theory: An Intermediate Text. South-Western Pub. Co., US

Gallant, A.P., Rossi, P.E., and G. Tauchen, (1992). Stock Prices and Volume, Rev. Financial Studies, 5 (2), 199-242

Ito, T. and W-L.Lin, 1993. Price Volatility And Volume. Spillovers Between The Tokyo And New York Stock Markets, NBER, WP 4592, 1-33

Karpoff, J.M., (1987). The Relation Between Price Changes and Trading Volume: A Survey The Journal of Financial and Quantitative Analysis, 22 (1), 109-126

Mankiw, N.G., Romer, D. and M.D. Shapiro, 1991. Stock Market Forecastability and

Volatility: A Statistical Appraisal, Rev.Economic Studies, 58,455-477
Markowitz, H. (1952). Portfolio Selection, J. Finance, 7(1), 77-91
Miloudi, A., Bouattour, M., Benkraiem, R., (2016). Relationships between Trading Volume, Stock Returns and Volatility: Evidence from the French Stock Market, Bankers, Markets & Investors, 144, 1-15
Muth, J.F., 1961. Rational Expectations and the Theory of Price Movements, Econometrica, 29, (3) 315-335.
Nakamura, E., and J. Steinsson, 2008. Five Facts About Prices: A Reevaluation Of Menu Cost Models. The Quarterly Jour. of Economics, 1415-1464
Olkhov, V. (2021a). Three Remarks On Asset Pricing, SSRN WP3852261, 1-20
Olkhov, V., (2021b). Theoretical Economics and the Second-Order Economic Theory. What is it?, MPRA WP 110893, 1-12
Olkhov, V. (2022). The Market-Based Asset Price Probability, MPRA WP115382, 1-18
Olkhov, V.(2023a). The Market-Based Probability of Stock Returns, SSRN WP4350975, 1-17
Olkhov, V., (2023b). Economic Complexity Limits Accuracy of Price Probability Predictions by Gaussian Distributions, SSRN WP 4550635, 1-23
Olkhov, V., (2023c). Theoretical Economics as Successive Approximations of Statistical Moments, SSRN WP 4586945, 1-17
Plerou, V., Gopikrishnan, P., Gabaix, X., Amaral, L.A., and H. E. Stanley, 2001. Price fluctuations, market activity and trading volume, Quantitative Finance, 1, 262–269
Poon, S-H. and C.W.J. Granger, (2003). Forecasting Volatility in Financial Markets: A Review, J. of Economic Literature, 41, 478–539
Shiryaev, A.N., (1999). Essentials of Stochastic Finance: Facts, Models, Theory, World Scientific Pub.Co., Singapore
Shreve, S. E., (2004). Stochastic calculus for finance, Springer finance series, NY, USA
Stigler, G.J., and J.K. Kindahl, 1970. The Dispersion of Price Movements, NBER, 88 - 94 in Ed. Stigler,G.J., and J.K. Kindahl ,The Behavior of Industrial Prices
Tauchen, G.E. and M. Pitts, 1983. The Price Variability-Volume Relationship On Speculative Markets, Econometrica, 51, (2), 485-505
Weyl, E.G., 2019. Price Theory, AEA J. of Economic Literature, 57(2), 329–384